\documentclass[twoside,twocolumn,9pt]{article}
\usepackage{extsizes}
\usepackage[super,sort&compress,comma]{natbib} 
\usepackage[version=3]{mhchem}
\usepackage[left=1.5cm, right=1.5cm, top=1.785cm, bottom=2.0cm]{geometry}
\usepackage{balance}
\usepackage{mathptmx}
\usepackage{sectsty}
\usepackage{graphicx} 
\usepackage{lastpage}
\usepackage[format=plain,justification=justified,singlelinecheck=false,font={stretch=1.125,small,sf},labelfont=bf,labelsep=space]{caption}
\usepackage{float}
\usepackage{fancyhdr}
\usepackage{fnpos}
\usepackage[english]{babel}
\addto{\captionsenglish}{%
  \renewcommand{\refname}{Notes and references}
}
\usepackage{array}
\usepackage{droidsans}
\usepackage{charter}
\usepackage[T1]{fontenc}
\usepackage[usenames,dvipsnames]{xcolor}
\usepackage{setspace}
\usepackage[compact]{titlesec}
\usepackage{hyperref}

\usepackage{epstopdf}

\usepackage{bm}

\definecolor{cream}{RGB}{222,217,201}

\begin{document}

\pagestyle{fancy}
\thispagestyle{plain}
\fancypagestyle{plain}{
\renewcommand{\headrulewidth}{0pt}
}

\makeFNbottom
\makeatletter
\renewcommand\LARGE{\@setfontsize\LARGE{15pt}{17}}
\renewcommand\Large{\@setfontsize\Large{12pt}{14}}
\renewcommand\large{\@setfontsize\large{10pt}{12}}
\renewcommand\footnotesize{\@setfontsize\footnotesize{7pt}{10}}
\makeatother

\renewcommand{\thefootnote}{\fnsymbol{footnote}}
\renewcommand\footnoterule{\vspace*{1pt}%
\color{cream}\hrule width 3.5in height 0.4pt \color{black}\vspace*{5pt}} 
\setcounter{secnumdepth}{5}

\makeatletter 
\renewcommand\@biblabel[1]{#1}            
\renewcommand\@makefntext[1]%
{\noindent\makebox[0pt][r]{\@thefnmark\,}#1}
\makeatother 
\renewcommand{\figurename}{\small{Fig.}~}
\sectionfont{\sffamily\Large}
\subsectionfont{\normalsize}
\subsubsectionfont{\bf}
\setstretch{1.125} 
\setlength{\skip\footins}{0.8cm}
\setlength{\footnotesep}{0.25cm}
\setlength{\jot}{10pt}
\titlespacing*{\section}{0pt}{4pt}{4pt}
\titlespacing*{\subsection}{0pt}{15pt}{1pt}

\fancyfoot{}
\fancyfoot[LO,RE]{\vspace{-7.1pt}\includegraphics[height=9pt]{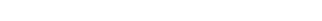}}
\fancyfoot[CO]{\vspace{-7.1pt}\hspace{13.2cm}\includegraphics{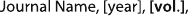}}
\fancyfoot[CE]{\vspace{-7.2pt}\hspace{-14.2cm}\includegraphics{head_foot/RF}}
\fancyfoot[RO]{\footnotesize{\sffamily{1--\pageref{LastPage} ~\textbar  \hspace{2pt}\thepage}}}
\fancyfoot[LE]{\footnotesize{\sffamily{\thepage~\textbar\hspace{3.45cm} 1--\pageref{LastPage}}}}
\fancyhead{}
\renewcommand{\headrulewidth}{0pt} 
\renewcommand{\footrulewidth}{0pt}
\setlength{\arrayrulewidth}{1pt}
\setlength{\columnsep}{6.5mm}
\setlength\bibsep{1pt}

\makeatletter 
\newlength{\figrulesep} 
\setlength{\figrulesep}{0.5\textfloatsep} 

\newcommand{\topfigrule}{\vspace*{-1pt}%
\noindent{\color{cream}\rule[-\figrulesep]{\columnwidth}{1.5pt}} }

\newcommand{\botfigrule}{\vspace*{-2pt}%
\noindent{\color{cream}\rule[\figrulesep]{\columnwidth}{1.5pt}} }

\newcommand{\dblfigrule}{\vspace*{-1pt}%
\noindent{\color{cream}\rule[-\figrulesep]{\textwidth}{1.5pt}} }

\makeatother

\twocolumn[
  \begin{@twocolumnfalse}
{\includegraphics[height=30pt]{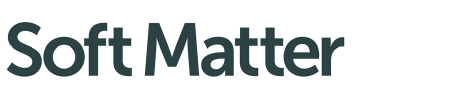}\hfill\raisebox{0pt}[0pt][0pt]{\includegraphics[height=55pt]{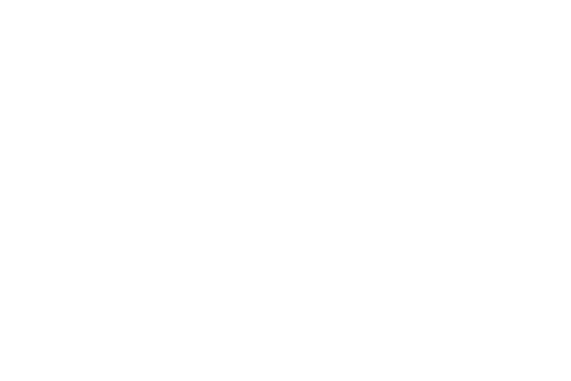}}\\[1ex]
\includegraphics[width=18.5cm]{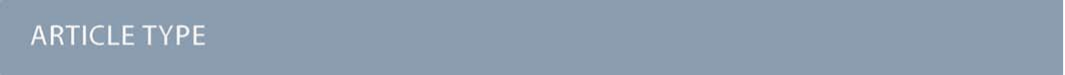}}\par
\vspace{1em}
\sffamily
\begin{tabular}{m{4.5cm} p{13.5cm} }

\includegraphics{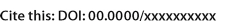} & \noindent\LARGE{\textbf{Compression-driven jamming in porous cohesive aggregates}} \\
\vspace{0.3cm} & \vspace{0.3cm} \\

 & \noindent\large{Sota Arakawa$^{\ast}$} \\

\includegraphics{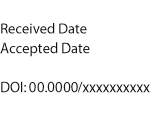} & \noindent\normalsize{

I investigate the compression-driven jamming behavior of two-dimensional porous aggregates composed of cohesive, frictionless disks.
Three types of initial aggregates are prepared using different aggregation procedures, namely, reaction-limited aggregation (RLA), ballistic particle--cluster aggregation (BPCA), and diffusion-limited aggregation (DLA), to elucidate the influence of aggregate morphology.
Using distinct-element-method simulations with a shrinking circular boundary, I numerically obtain the pressure as a function of the packing fraction $\phi$.
For the densest RLA and the intermediate BPCA aggregates, a clear jamming transition is observed at a critical packing fraction $\phi_{\rm J}$, below which the pressure vanishes and above which a finite pressure emerges; the transition is less distinct for the most porous DLA aggregates.
The jamming threshold depends on the initial structure and, when extrapolated to infinite system size, approaches $\phi_{\rm J} = 0.765 \pm 0.004$ for RLA, $0.727 \pm 0.004$ for BPCA, and $0.602 \pm 0.023$ for DLA, where the errors denote the standard error.
Above $\phi_{\rm J}$, the pressure follows $P \approx  A {( \phi - \phi_{\rm J} )}^{2}$, which implies that the bulk modulus $K$ of jammed aggregates is proportional to $\phi - \phi_{\rm J}$.
Rigid-cluster analysis of jammed aggregates shows that the average coordination number within the largest rigid cluster increases linearly with $\phi - \phi_{\rm J}$.
Taken together, these relations suggest that the elastic response of compressed porous aggregates is analogous to that of random spring networks.

} \\

\end{tabular}

 \end{@twocolumnfalse} \vspace{0.6cm}

  ]

\renewcommand*\rmdefault{bch}\normalfont\upshape
\rmfamily
\section*{}
\vspace{-1cm}


\footnotetext{\textit{Center for Mathematical Science and Advanced Technology, Japan Agency for Marine-Earth Science and Technology, 3173-25, Showa-machi, Kanazawa-ku, Yokohama 236-0001, Japan. E-mail: arakawas@jamstec.go.jp}}





\section{Introduction}

Soft-matter systems consisting of large numbers of discrete particles, such as granular materials, emulsions, foams, and cellular tissues, jam into mechanically rigid solids once their packing fraction, $\phi$, exceeds a critical value \citep{2010JPCM...22c3101V}.
This jamming transition has been explored extensively over the past few decades \citep{2000PhRvL..84.4160M, 2002PhRvL..88g5507O, 2003PhRvE..68a1306O, 2007PhRvL..98e8001M, 2009EL.....8734004E, 2010ARCMP...1..347L, 2010RvMP...82..789P}.
In athermal assemblies of frictionless particles with purely repulsive contacts, macroscopic observables, including pressure, bulk modulus, and shear modulus, display critical behavior in the vicinity of the jamming point $\phi_{\rm J}$ \citep{2003PhRvE..68a1306O, 2009EL.....8734004E, 2011PhRvB..83r4205Z, 2025JAP...137e0901Z}.

Although much of the foundational work has dealt with purely repulsive, frictionless systems, interparticle cohesion and friction are ubiquitous in real materials.
Granular packings that are frictional but otherwise repulsive jam at significantly lower $\phi$ than their frictionless counterparts \citep{2020PhRvE.102c2903S}.
The impact of attractive forces on the mechanical response has likewise been examined in a series of studies \citep{2008PhRvL.100b8001L, 2011PhRvE..84a1303L, 2018PhRvL.121r8002K, 2020PhRvR...2c2047K, 2025JPSJ...94h4801Y, 2024PhRvE.109c4406G}. 
For example, \citet{2020PhRvR...2c2047K} showed that the bulk modulus (and hence the pressure) of two-dimensional cohesive, frictionless packings follows a qualitatively different $\phi$-dependence from that of purely repulsive systems, a trend later confirmed in three-dimensional systems near jamming by \citet{2025JPSJ...94h4801Y}.

Cohesive forces do more than modify contact mechanics; they also drive aggregation before compression.
Under dilute conditions, attractive particles spontaneously form porous clusters.
Such aggregates are widespread in nature, ranging from colloidal gels to cosmic dust \citep{1996EnST...30.1911J, 2008ARA&A..46...21B}, and their morphology is dictated by the coagulation pathway \citep{1991RvGeo..29..317M}.
Most jamming studies of cohesive, frictionless particles, however, start from initially dispersed states and therefore overlook how the pre-compression structure influences the subsequent mechanical response.

Here, I numerically investigate compression-driven jamming in two-dimensional porous aggregates composed of cohesive, frictionless disks.
Three types of initial aggregates are prepared using different aggregation procedures, and I perform numerical compression tests using the distinct element method (DEM) \citep{doi:10.1680/geot.1979.29.1.47}.
By analyzing the pressure as a function of the packing fraction $\phi$, I determine the jamming threshold $\phi_{\rm J}$ and its dependence on the initial aggregate morphology.
I further identify a critical power-law relation between the pressure and $\phi - \phi_{\rm J}$ above $\phi_{\rm J}$ for each aggregate type.
These results indicate that the elastic response of compressed porous aggregates is analogous to that of random spring networks.

\section{Models}

\subsection{Particle Interaction}

I consider systems of monodisperse disk-shaped particles initially confined by a circular wall of radius $R_{\rm wall}$ and simulate quasi–static compression by gradually reducing $R_{\rm wall}$.
Each particle has unit mass, $m_{\rm disk} = 1$, and unit radius, $R_{\rm disk} = 1$.
The number of particles is set to $N = 1024$, $4096$, or $16384$ for each simulation.
The equation of motion is given by
\begin{equation}
m_{\rm disk} \frac{{\rm d}^{2}{\bm r}_{i}}{{\rm d}t^{2}} = {\bm F}_{i},
\end{equation}
where ${\bm r}_{i}$ is the position of particle $i$, and ${\bm F}_{i}$ is the total force acting upon it.
Rotational degrees of freedom are neglected in this study.
The timestep for numerical integration is fixed at ${\Delta t} = 0.02$.
For two particles $i$ and $j$ in contact, the interparticle force, ${\bm F}_{ij}$, is given by the sum of the elastic term ${\bm F}_{\rm e}$ and dissipative term ${\bm F}_{\rm d}$:
\begin{equation}
{\bm F}_{ij} = {\bm F}_{\rm e} + {\bm F}_{\rm d}.
\end{equation}
I consider a simple dissipation model that is widely used to describe energy dissipation in foam bubbles \citep{1997PhRvE..55.1739D}:
\begin{equation}
{\bm F}_{\rm d} = - \frac{C_{\rm d}}{2} {\left( {\bm v}_{i} - {\bm v}_{j} \right)},
\end{equation}
where ${\bm v}_{i}$ is the velocity of particle $i$, and the dissipation coefficient for damping ($C_{\rm d}$) is set to $C_{\rm d} = 1$.
In this study, ${\bm F}_{\rm d}$ was introduced to facilitate the energy and structural relaxation of the system.
Rheological properties, such as viscosity, are beyond the scope of this study, and I expect that the specific choice of ${\bm F}_{\rm d}$ does not significantly affect the main conclusions.

\begin{figure}
\centering
  \includegraphics[width=\columnwidth]{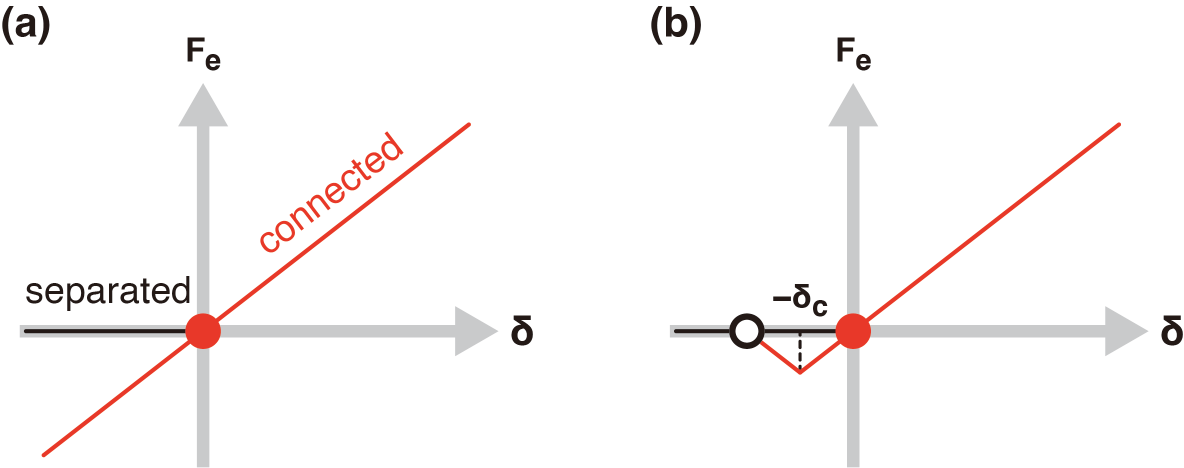}
  \caption{\
  Schematics of elastic interaction models.
  (a) Unbreakable contact model.
  $F_{\rm e} = 0$ when the two particles are separated (black line).
  A contact is formed when the particles collide (red filled circle).
  Once a contact has formed, $F_{\rm e}$ is given by Equation (\ref{eq:F_e}) (red line).
  A repulsive force acts when the compression length is positive, whereas an attractive force acts when it is negative.
  In this model, a contact never breaks once it has formed.
  (b) Breakable contact model.
  $F_{\rm e} = 0$ when the two particles are separated (black line).
  A contact is formed when the particles collide (red filled circle).
  Once a contact has formed, $F_{\rm e}$ is given by Equation (\ref{eq:F_e_break}) (red line).
  In this model, the contact breaks when $\delta$ reaches $- 2 \delta_{\rm c}$ (black open circle).
  }
  \label{fig:1}
\end{figure}

The elastic term ${\bm F}_{\rm e}$ is a function of the compression length (i.e., penetration depth) between two particles, $\delta_{ij} = 2 R_{\rm disk} - {| {\bm r}_{i} - {\bm r}_{j} |}$.
I assume that a contact between particles $i$ and $j$ forms when they collide ($\delta_{ij} = 0$).
A repulsive force acts when the compression length is positive, whereas an attractive force acts when it is negative.
In this study, I consider two types of elastic interaction, namely, the unbreakable and the breakable contact models (Figure \ref{fig:1}).
For the unbreakable contact model, I do not allow a contact to break once it has formed, and ${\bm F}_{\rm e}$ is given by
\begin{equation}
{\bm F}_{\rm e} = \frac{k}{2} \delta_{ij} {\bm n}_{ij},
\label{eq:F_e}
\end{equation}
where $\delta_{ij} = 2 R_{\rm disk} - {| {\bm r}_{i} - {\bm r}_{j} |}$ denotes the compression length (i.e., penetration depth) between two particles, and ${\bm n}_{ij} = {( {\bm r}_{i} - {\bm r}_{j} )} / {| {\bm r}_{i} - {\bm r}_{j} |}$ is the normal unit vector.
The spring constant ($k$) is set to $k = 10$.

In contrast, for the breakable contact model, I consider the weakening and eventual breakage of contacts when $\delta_{ij}$ reaches prescribed threshold values.
In this study, I assume a simple model:
\begin{equation}
{\bm F}_{\rm e} = 
\begin{cases}
{( k / 2 )} \delta_{ij} {\bm n}_{ij}                             &  ( \delta_{ij} \ge - \delta_{\rm c} ), \\
{( k / 2 )} {( - 2 \delta_{\rm c} - \delta_{ij} )} {\bm n}_{ij}  &  ( - 2 \delta_{\rm c} < \delta_{ij} < - \delta_{\rm c} ),
\end{cases}
\label{eq:F_e_break}
\end{equation}
and a contact breaks when $\delta_{ij} = - 2 \delta_{\rm c}$.
The critical length $\delta_{\rm c}$ is set to $\delta_{\rm c} = 0.01$.

The force exerted on particle $i$ interacting with the wall, ${\bm F}_{\rm wall}$, is
\begin{eqnarray}
{\bm F}_{\rm wall}    & = & {\bm F}_{\rm e, wall} + {\bm F}_{\rm d, wall}, \\
{\bm F}_{\rm d, wall} & = & - C_{\rm d} {\left( {\bm v}_{i} - {\bm v}_{\rm wall} \right)},
\label{eq:Fwall}
\end{eqnarray}
where ${\bm v}_{\rm wall} = v_{\rm wall} {\bm n}_{\rm wall}$ is the velocity of the wall at the contact point.
The speed of the wall, $v_{\rm wall}$, is defined as $v_{\rm wall} = - {\rm d}R_{\rm wall} / {\rm d}t$.
For the unbreakable contact model, ${\bm F}_{\rm e, wall}$ is given by
\begin{equation}
{\bm F}_{\rm e, wall} = k \delta_{\rm wall} {\bm n}_{\rm wall},
\label{eq:F_e_wall}
\end{equation}
where $\delta_{\rm wall} = R_{\rm disk} - {( R_{\rm wall} - {| {\bm r}_{i} |} )}$ is the compression length with wall, and ${\bm n}_{\rm wall} = - {\bm r}_{i} / {| {\bm r}_{i} |}$ is the normal unit vector.
For the breakable contact model, ${\bm F}_{\rm e, wall}$ is given by
\begin{equation}
{\bm F}_{\rm e, wall} = 
\begin{cases}
k \delta_{\rm wall} {\bm n}_{\rm wall}                           &  ( \delta_{\rm wall} \ge - 0.5 \delta_{\rm c} ), \\
k {( - \delta_{\rm c} - \delta_{\rm wall} )} {\bm n}_{\rm wall}  &  ( - \delta_{\rm c} < \delta_{\rm wall} < - 0.5 \delta_{\rm c} ),
\end{cases}
\end{equation}
and a contact breaks when $\delta_{\rm wall} = - \delta_{\rm c}$.
The packing fraction of the system is $\phi = N / {R_{\rm wall}}^{2}$, and the pressure, $P$, is defined as follows:
\begin{equation}
P = \frac{1}{2 \pi R_{\rm wall}} \sum {\bm F}_{\rm wall} \cdot {\bm n}_{\rm wall},
\end{equation}
where the summation is taken over all particle--wall contacts.

In the present study, I perform numerical simulations using the unbreakable contact model to obtain the main results (Section \ref{sec:results}).
The impact of contact breakage is discussed in Section \ref{sec:break}.

\subsection{Initial Aggregate Structure}

\begin{figure}
\centering
  \includegraphics[width=\columnwidth]{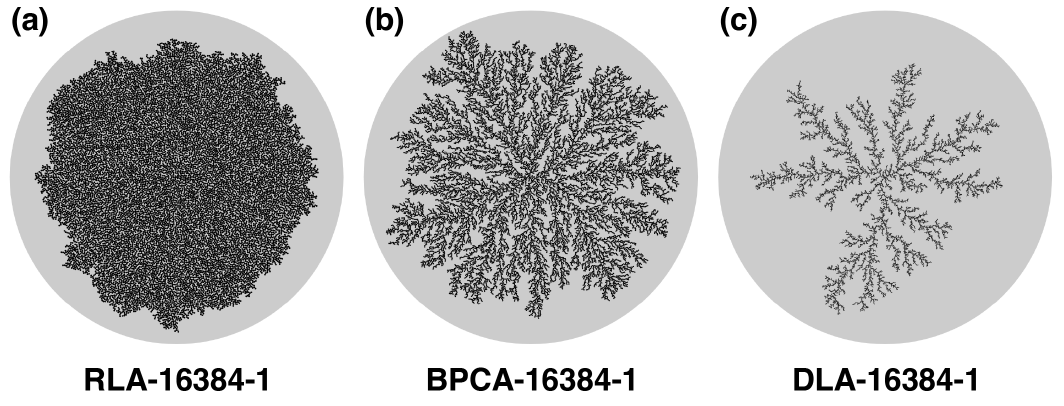}
  \caption{\
  Three types of aggregates prepared using distinct aggregation procedures.
  Panels (a), (b), and (c) show the initial structures of DLA, BPCA, and RLA aggregates with $N = 16384$, respectively.
  }
  \label{fig:2}
\end{figure}

I employ three types of aggregates prepared via distinct procedures as initial configurations: reaction-limited aggregation (RLA), ballistic particle--cluster aggregation (BPCA), and diffusion-limited aggregation (DLA).
These aggregates are generated in continuous (off-lattice) space.
The corresponding structures are illustrated in Figures \ref{fig:2}(a), \ref{fig:2}(b), and \ref{fig:2}(c), respectively.
Below, I briefly describe these aggregation protocols (see also Ref.\citep{meakin1999historical}).

\subsubsection{Reaction-Limited Aggregation (RLA)}

RLA, also known as the Eden growth model, is a sequential growth process in which particles attach to randomly selected sites on a pre-existing cluster.
RLA-type cluster formation typically occurs when the probability of particle attachment upon collision is low, and the overall growth is limited by the attachment kinetics.
Such structures are widely observed in nature, ranging from bacterial colonies \citep{2025PhRvE.111a4406K} to topological-defect turbulence in nematic liquid crystals \citep{2012JSP...147..853T}.  
The structural properties of RLA clusters have been extensively investigated \citep{2012JSMTE..05..007T, 2018JPSJ...87a4005K}.  
An initial RLA aggregate with $N = 16384$ is shown in Figure \ref{fig:2}(a), and the initial wall radius is set to $R_{\rm wall, ini} = 200$.

\subsubsection{Ballistic Particle--Cluster Aggregation (BPCA)}

BPCA is a sequential growth process that occurs via collisions between a particle and a cluster.  
In this process, a particle launched from a distant point moves ballistically and adheres to the surface of a growing cluster upon collision.
BPCA-type cluster formation typically arises in dilute media where the mean free path of particles exceeds the aggregate size.
BPCA has been widely used as a model for cosmic dust aggregates \citep{2009ApJ...702.1490W, 2023A&A...670L..21A, 2024MNRAS.534.2125K} and atmospheric aerosols \citep{1990AerST..12..876S}.
An initial BPCA aggregate with $N = 16384$ is shown in Figure \ref{fig:2}(b), with the initial wall radius set to $R_{\rm wall, ini} = 270$.

\subsubsection{Diffusion-Limited Aggregation (DLA)}

DLA is a sequential growth process driven by collisions between a particle and a cluster.
In this process, a randomly walking particle starting far from the origin collides with a growing cluster and adheres to its surface.
DLA-type cluster formation is typically observed when growth occurs within a diffusion field \citep{1981PhRvL..47.1400W}.
Such structures are widely observed in nature, ranging from bacterial colonies \citep{1990PhyA..168..498M} to dendritic crystals formed via electrochemical deposition \citep{1986PhRvL..56.1264G}.
The structural properties of DLA clusters have been extensively investigated \citep{2009JPSJ...78a3605O, 2024JPSJ...93b4401A}.
An initial DLA aggregate with $N = 16384$ is shown in Figure \ref{fig:2}(c), with the initial wall radius set to $R_{\rm wall, ini} = 600$.

\subsection{Compression and Relaxation}

In this study, I obtain statically compressed aggregates through a two-step numerical procedure described below (Figure \ref{fig:3}).
In the first stage, the wall radius $R_{\rm wall}$ is continuously decreased, producing a series of configurations ${\bm C}$ with packing fractions $\phi$ (Figures \ref{fig:3}(b) and \ref{fig:3}(c)).
In the second stage, the generated configurations ${\bm C} ( \phi )$ are relaxed at constant $R_{\rm wall}$ to new configurations ${\bm C}_{\rm relax} ( \phi )$ (Figures \ref{fig:3}(d) and \ref{fig:3}(e)).
The pressure $P = P ( \phi )$ is then measured in the final static configuration.

\begin{figure}
\centering
  \includegraphics[width=0.8\columnwidth]{Fig3.pdf}
  \caption{\
  Snapshots of a compressed aggregate obtained using a two-step numerical procedure.
  (a) Initial configuration of {\sf BPCA-16384-1}.
  (b), (c) Configurations at $\phi = 0.694$ and $0.717$ during the first compression stage.
  (d), (e) Configurations at $\phi = 0.694$ and $0.717$ after relaxation.
  The red particles in panels (d) and (e) form the largest rigid cluster (see Section \ref{sec:rigid}).
  }
  \label{fig:3}
\end{figure}

During the first compression stage, the wall radius $R_{\rm wall}$ decreases exponentially with time according to a characteristic timescale $\tau_{\rm comp}$:
\begin{equation}
R_{\rm wall} {( t )} = R_{\rm wall, ini} \exp{\left( - t / \tau_{\rm comp} \right)},
\label{eq:R_wall}
\end{equation}
where $R_{\rm wall, ini}$ is the initial wall radius, and I set $\tau_{\rm comp} = 10^{6}$.
In the second relaxation stage, the wall radius is held constant ($R_{\rm wall} = \mathrm{const.}$), and the aggregate is allowed to relax over a waiting time of $\tau_{\rm relax} = 2 \times 10^{5}$.

\section{Results}
\label{sec:results}

In this study, I prepare three types of initial aggregates (RLA, BPCA, and DLA) for three aggregate sizes ($N = 16384$, $4096$, and $1024$).
For each combination of type and size, I generate five aggregates with different initial configurations.
Each simulation run is labeled, for example, as {\sf RLA-16384-1} or {\sf DLA-1024-5}.

\subsection{Pressure and Jamming Point}
\label{sec:relax}

\begin{figure*}
\centering
  \includegraphics[width=\textwidth]{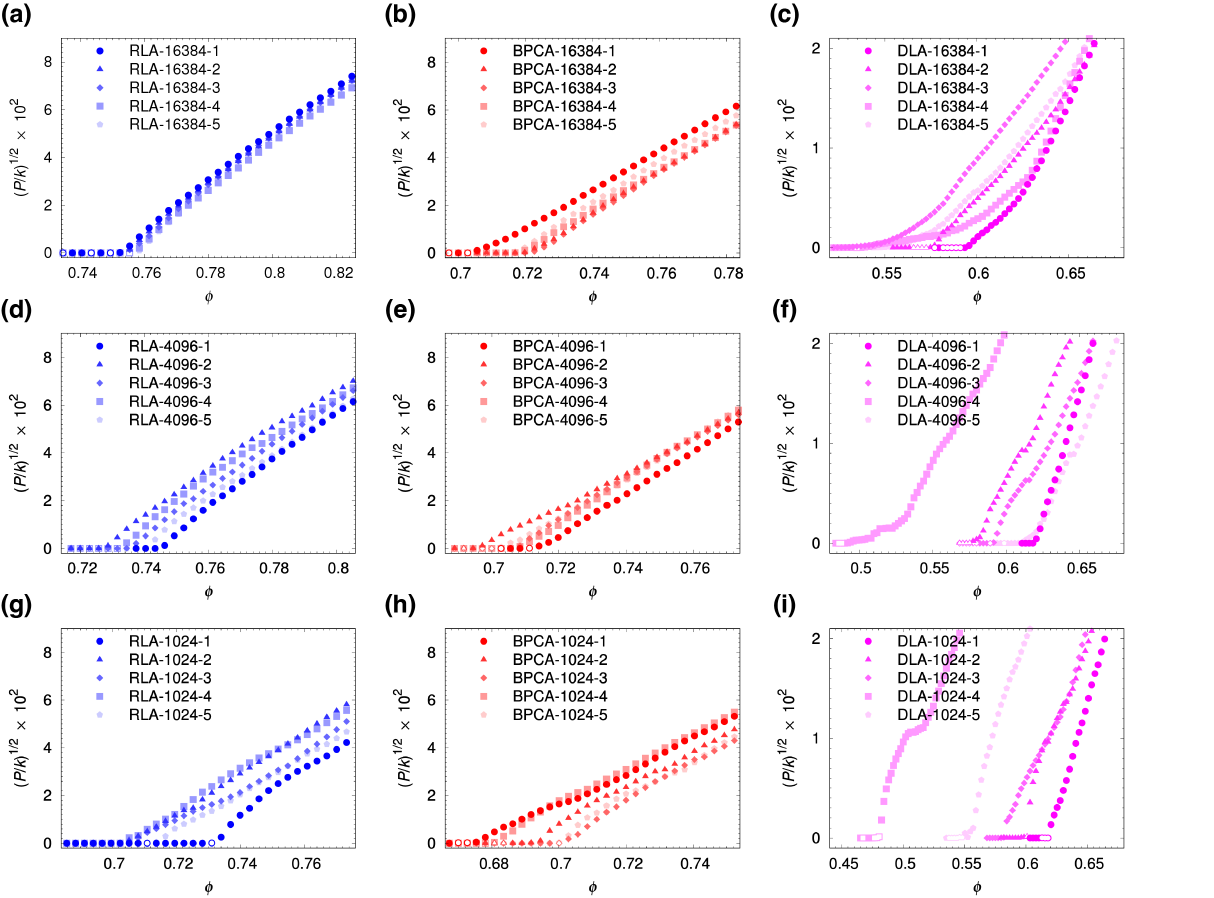}
  \caption{\
  Normalized pressure $P/k$ after relaxation.
  The figure shows the relationship between ${\left( P / k \right)}^{1/2}$ and $\phi$ near the jamming transition.
  (a)--(c) Results for $N = 16384$.
  (d)--(f) Results for $N = 4096$.
  (g)--(i) Results for $N = 1024$.
  Data with positive $P$ are plotted using filled symbols, whereas data with negative $P$ are plotted as $- P$ using open symbols.
  }
  \label{fig:4}
\end{figure*}

I analyze the pressure as a function of the packing fraction.
As $P$ is proportional to $k$ in static configurations, I consider the normalized pressure $P/k$ in this study.
Figure \ref{fig:4} shows the relationship between ${\left( P/k \right)}^{1/2}$ and $\phi$ near the jamming transition.
I find that there exists a critical packing fraction $\phi_{\rm J}$ such that the pressure vanishes for $\phi < \phi_{\rm J}$ and increases with $\phi$ for $\phi > \phi_{\rm J}$.
The value of $\phi_{\rm J}$ depends on the initial aggregate configuration.

Figures \ref{fig:4}(a), \ref{fig:4}(b), and \ref{fig:4}(c) show the results for RLA, BPCA, and DLA aggregates with $N = 16384$, respectively.
For RLA and BPCA, I find a linear relationship between $\left( P / k \right)^{1/2}$ and $\phi$ for $\phi > \phi_{\rm J}$.
These results indicate that $P / k$ is approximately proportional to ${\left( \phi - \phi_{\rm J} \right)}^{2}$ in these cases.
Results for $N = 4096$ and $1024$ are also shown in Figures \ref{fig:4}(d)--\ref{fig:4}(f) and \ref{fig:4}(g)--\ref{fig:4}(i), respectively.
The linear relationship between $\left( P / k \right)^{1/2}$ and $\phi$ is less clear for some DLA aggregates (e.g., {\sf DLA-16384-4}), although other DLA aggregates (e.g., {\sf DLA-16384-2} and {\sf DLA-4096-1}) do exhibit an approximately linear trend.

I estimate the value of $\phi_{\rm J}$ for each aggregate by least-squares fitting, assuming the following critical scaling relation with exponent $\alpha = 2$:
\begin{equation}
P / k =
\begin{cases}
A \left( \phi - \phi_{\rm J} \right)^{\alpha}   & {\left( \phi > \phi_{\rm J} \right)}, \\
0                                               & {\left( \phi \le \phi_{\rm J} \right)}. \\
\end{cases}
\label{eq:p/k}
\end{equation}
The least-squares fits are performed using data in the range $0.005 < {( P / k )}^{1/2} < 0.04$ for RLA and BPCA aggregates, and $0.005 < {( P / k )}^{1/2} < 0.02$ for DLA aggregates.
The fitting parameters for all runs are summarized in Table \ref{table:1}.
I acknowledge that, in the immediate vicinity of $\phi_{\rm J}$, i.e., in the
region where ${( P/k )}^{1/2} < 0.005$, the values of $P/k$ exhibit considerable scatter, and thus there is a limit to how close to $\phi_{\rm J}$ I can take the fitting range.

\begin{table}
\small
  \caption{\
  Fitting parameters ($\phi_{\rm J}$ and $A$) for the packing-fraction dependence of the normalized pressure $P / k$ (Equation (\ref{eq:p/k})).
  The least-squares fits are performed using data in the range $0.005 < {( P / k )}^{1/2} < 0.04$ for RLA and BPCA aggregates, and $0.005 < {( P / k )}^{1/2} < 0.02$ for DLA aggregates.
  }
  \begin{tabular*}{0.48\textwidth}{@{\extracolsep{\fill}}lcc}
    \hline
    Aggregate ID  & $\phi_{\rm J}$ & $A$ \\
    \hline
    
    {\sf RLA-16384-1}   & $0.7516$  & $1.173$  \\
    {\sf RLA-16384-2}   & $0.7526$  & $1.121$  \\
    {\sf RLA-16384-3}   & $0.7546$  & $1.258$  \\
    {\sf RLA-16384-4}   & $0.7550$  & $1.095$  \\
    {\sf RLA-16384-5}   & $0.7561$  & $1.222$  \\ \\

    {\sf RLA-4096-1}    & $0.7435$  & $1.068$  \\
    {\sf RLA-4096-2}    & $0.7261$  & $0.951$  \\
    {\sf RLA-4096-3}    & $0.7339$  & $0.915$  \\
    {\sf RLA-4096-4}    & $0.7305$  & $1.022$  \\
    {\sf RLA-4096-5}    & $0.7368$  & $0.821$  \\ \\
    
    {\sf RLA-1024-1}    & $0.7266$  & $0.871$  \\
    {\sf RLA-1024-2}    & $0.7064$  & $0.738$  \\
    {\sf RLA-1024-3}    & $0.7060$  & $0.444$  \\
    {\sf RLA-1024-4}    & $0.7041$  & $0.736$  \\
    {\sf RLA-1024-5}    & $0.7128$  & $0.588$  \\ \\
    
    {\sf BPCA-16384-1}  & $0.7074$  & $0.659$  \\
    {\sf BPCA-16384-2}  & $0.7218$  & $0.819$  \\
    {\sf BPCA-16384-3}  & $0.7226$  & $0.820$  \\
    {\sf BPCA-16384-4}  & $0.7185$  & $0.708$  \\
    {\sf BPCA-16384-5}  & $0.7163$  & $0.789$  \\ \\
    
    {\sf BPCA-4096-1}   & $0.7138$  & $0.756$  \\
    {\sf BPCA-4096-2}   & $0.6953$  & $0.481$  \\
    {\sf BPCA-4096-3}   & $0.7058$  & $0.737$  \\
    {\sf BPCA-4096-4}   & $0.7091$  & $0.886$  \\
    {\sf BPCA-4096-5}   & $0.7044$  & $0.660$  \\ \\
    
    {\sf BPCA-1024-1}   & $0.6739$  & $0.398$  \\
    {\sf BPCA-1024-2}   & $0.6885$  & $0.548$  \\
    {\sf BPCA-1024-3}   & $0.6987$  & $0.698$  \\
    {\sf BPCA-1024-4}   & $0.6774$  & $0.542$  \\
    {\sf BPCA-1024-5}   & $0.6937$  & $0.544$  \\ \\

    {\sf DLA-16384-1}   & $0.6119$  & $0.148$  \\
    {\sf DLA-16384-2}   & $0.5824$  & $0.055$  \\
    {\sf DLA-16384-3}   & $0.5630$  & $0.057$  \\
    {\sf DLA-16384-4}   & $0.6054$  & $0.126$  \\
    {\sf DLA-16384-5}   & $0.5773$  & $0.055$  \\ \\

    {\sf DLA-4096-1}    & $0.6207$  & $0.261$  \\
    {\sf DLA-4096-2}    & $0.5814$  & $0.101$  \\
    {\sf DLA-4096-3}    & $0.5938$  & $0.086$  \\
    {\sf DLA-4096-4}    & $0.5183$  & $0.063$  \\
    {\sf DLA-4096-5}    & $0.6186$  & $0.119$  \\ \\

    {\sf DLA-1024-1}    & $0.6201$  & $0.204$  \\
    {\sf DLA-1024-2}    & $0.5846$  & $0.084$  \\
    {\sf DLA-1024-3}    & $0.5803$  & $0.074$  \\
    {\sf DLA-1024-4}    & $0.4607$  & $0.050$  \\
    {\sf DLA-1024-5}    & $0.5465$  & $0.148$  \\
    
    \hline
  \end{tabular*}
  \label{table:1}
\end{table}

\begin{figure}
\centering
  \includegraphics[width=\columnwidth]{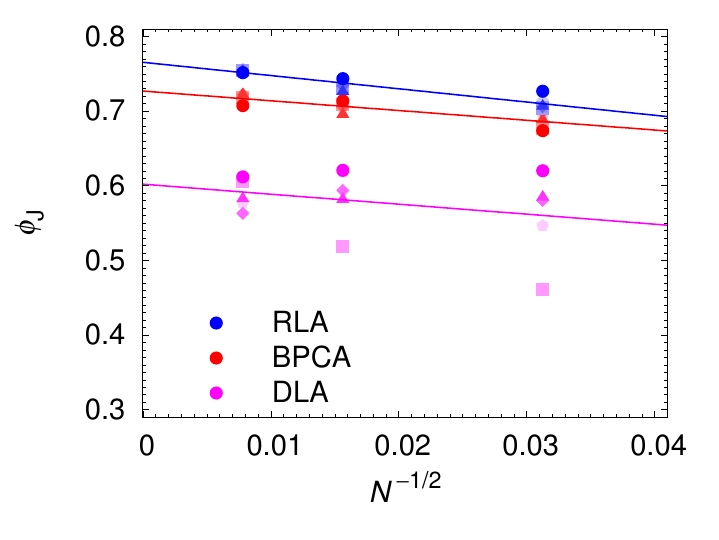}
  \caption{\
  System-size dependence of the jamming point $\phi_{\rm J}$.
  Symbol shapes (circle, triangle, diamond, square, and pentagon) correspond to those in Figure \ref{fig:4}.
  The solid lines represent the best fits obtained by least-squares fitting to Equation (\ref{eq:phi_J_fit}).
  }
  \label{fig:5}
\end{figure}

To elucidate the effect of finite system size $N$ on the jamming point $\phi_{\rm J}$, I investigate how $\phi_{\rm J}$ depends on $N$ for the three types of aggregates.
Figure \ref{fig:5} shows the system-size dependence of $\phi_{\rm J}$.
I find that $\phi_{\rm J}$ is approximately given by the following empirical relation:
\begin{equation}
\phi_{\rm J} = \phi_{{\rm J}, \infty} - c N^{-1/2},
\label{eq:phi_J_fit}
\end{equation}
where $\phi_{{\rm J}, \infty}$ is the jamming point for an infinitely large aggregate and $c$ is a fitting constant.
From the fitting shown in Figure \ref{fig:5}, I obtain $\phi_{{\rm J}, \infty} = 0.765 \pm 0.004$ for RLA, $0.727 \pm 0.004$ for BPCA, and $0.602 \pm 0.023$ for DLA, where the quoted errors denote the standard errors.
These results highlight that the initial aggregate morphology strongly influences the jamming point and underscore the importance of the pre-compression aggregation history in cohesive granular materials.

I briefly discuss the bulk modulus of the compressed aggregates.
The bulk modulus $K$ is defined as
\begin{equation}
K = \phi \frac{{\rm d}P}{{\rm d}\phi}.
\end{equation}
Given that the pressure follows the critical scaling described by Equation (\ref{eq:p/k}), the bulk modulus is expressed as
\begin{equation}
K / k =
\begin{cases}
2 A \phi {\left( \phi - \phi_{\rm J} \right)}  & {\left( \phi > \phi_{\rm J} \right)}, \\ 
0                                              & {\left( \phi \le \phi_{\rm J} \right)}. \\ 
\end{cases}
\label{eq:bulk}
\end{equation}
Equation (\ref{eq:bulk}) implies that $K$ varies continuously across $\phi_{\rm J}$.

The dependence of pressure on the filling factor in cohesive granular materials differs qualitatively from that in purely repulsive granular systems, where ${\bm F}_{\rm e}$ is defined as
\begin{equation}
{\bm F}_{\rm e} = 
\begin{cases}
{( k / 2 )} \delta_{ij} {\bm n}_{ij}   & ( \delta_{ij} \ge 0 ), \\
0                                      & ( \delta_{ij} < 0 ).
\end{cases}
\end{equation}
The elastic properties of granular systems composed of frictionless particles with linear repulsive interactions have been extensively studied in previous works \citep{2002PhRvL..88g5507O, 2003PhRvE..68a1306O}.
It is known that the pressure scales as $P/k = A {( \phi - \phi_{\rm J} )}$ for $\phi > \phi_{\rm J}$ \citep{2003PhRvE..68a1306O}, resulting in a discontinuous jump of the bulk modulus $K$ at $\phi_{\rm J}$.
For granular systems composed of frictionless Hertzian repulsive particles,
one instead finds $P/k = A {( \phi - \phi_{\rm J} )}^{3/2}$ for $\phi > \phi_{\rm J}$ \citep{2003PhRvE..68a1306O}.
In Section \ref{sec:rigid}, I perform a rigid-cluster analysis to understand the origin of this difference.

In my simulations, a linear relationship between ${( P / k )}^{1/2}$ and $\phi - \phi_{\rm J}$ holds with high accuracy for RLA and BPCA with $N = 16384$ (Figures \ref{fig:4}(a) and \ref{fig:4}(b)).
Therefore, in the analyses presented in Sections \ref{sec:rigid} and \ref{sec:delta}, I focus on these results.

\subsection{Rigid Cluster Analysis}
\label{sec:rigid}

I perform a rigid-cluster analysis to identify the largest rigid cluster in each final static configuration.
Rigid clusters are defined as those whose zero-frequency vibrational modes correspond only to rigid-body motions, and I use the pebble-game algorithm \citep{1995PhRvL..75.4051J} to identify them.
The red particles in Figures \ref{fig:3}(d) and \ref{fig:3}(e) form the largest rigid cluster.
An aggregate confined by a boundary wall enters a jammed state when a rigid cluster has more than two contact points with the boundary wall.
This rigidity percolation is evident in Figure \ref{fig:3}(e); the largest rigid cluster has many contacts with the wall for $\phi > \phi_{\rm J}$.

\begin{figure*}
\centering
  \includegraphics[width=0.8\textwidth]{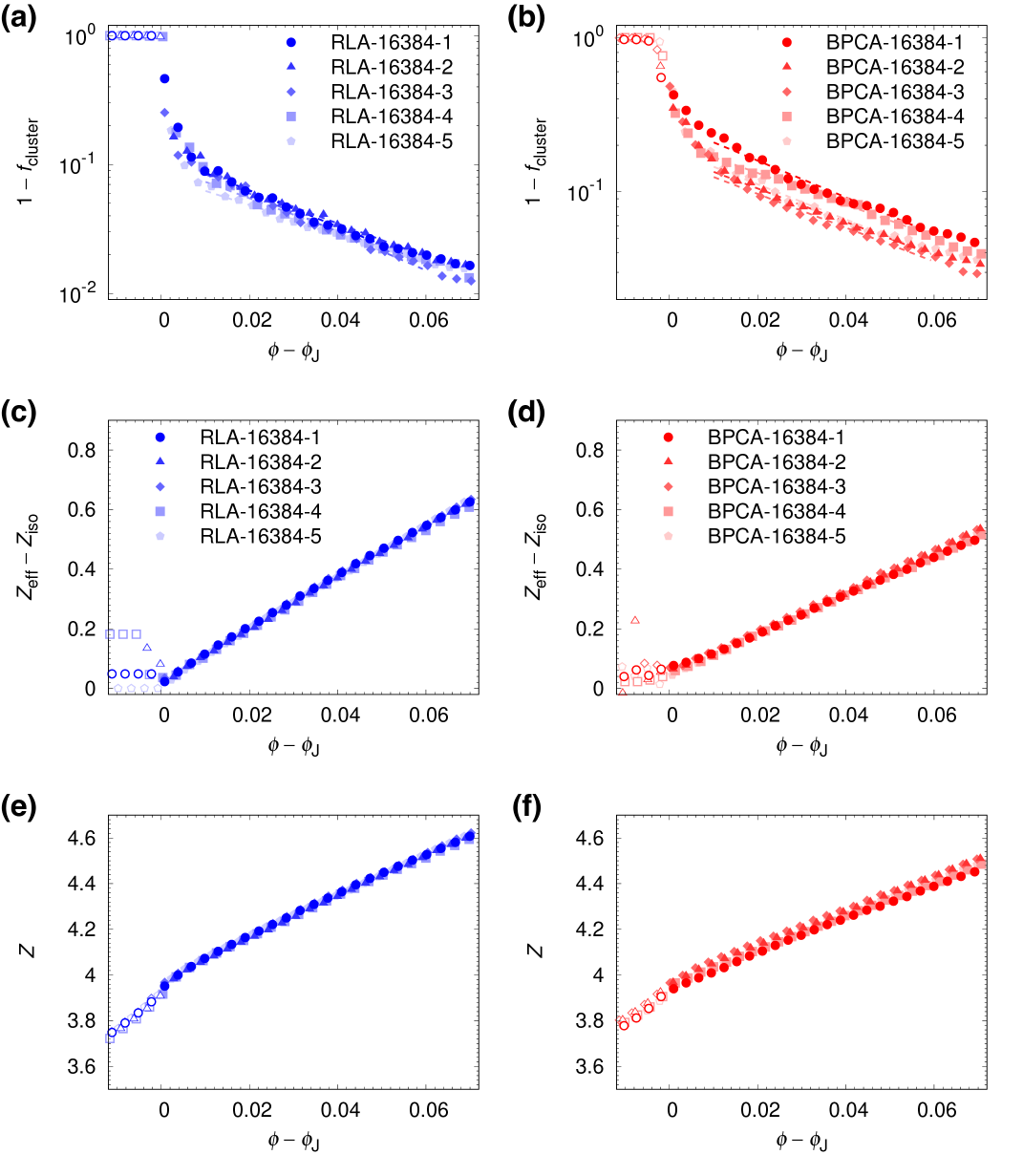}
  \caption{\
  (a), (b) Fraction of the largest rigid cluster, $f_{\rm cluster} = N_{\rm cluster} / N$, as a function of $\phi - \phi_{\rm J}$.
  Panels (a) and (b) show the results for the RLA and BPCA aggregates with $N = 16384$, respectively.
  Symbol shapes (circle, triangle, diamond, square, and pentagon) correspond to those in Figure \ref{fig:4}.
  Filled symbols indicate data for $\phi \ge \phi_{\rm J}$, whereas open symbols indicate data for $\phi < \phi_{\rm J}$.
  The dashed lines represent the fits obtained using data in the range $0.01 < \phi - \phi_{\rm J} < 0.06$ (Equation (\ref{eq:f})).
  (c), (d)  Coordination number within the largest rigid cluster, $Z_{\rm eff}$, as a function of $\phi - \phi_{\rm J}$.
  Panels (c) and (d) show the results for the RLA and BPCA aggregates with $N = 16384$, respectively.
  (e), (f)  Coordination number of the whole aggregate, $Z$, as a function of $\phi - \phi_{\rm J}$.
  Panels (e) and (f) show the results for the RLA and BPCA aggregates with $N = 16384$, respectively.
  }
  \label{fig:6}
\end{figure*}

I define $N_{\rm cluster}$ as the number of particles that constitute the largest rigid cluster, and the fraction of the largest rigid cluster as $f_{\rm cluster} = N_{\rm cluster} / N$.
Figures \ref{fig:6}(a) and \ref{fig:6}(b) show $1 - f_{\rm cluster}$ as a function of $\phi - \phi_{\rm J}$.
Results for the RLA and BPCA aggregates with $N = 16384$ are shown in Figures \ref{fig:6}(a) and \ref{fig:6}(b), respectively.
I find that $1 - f_{\rm cluster}$ decreases sharply at $\phi_{\rm J}$ in both cases, and that it can be approximated by
\begin{equation}
1 - f_{\rm cluster} = f_{0} \exp{\left( - \frac{\phi - \phi_{\rm J}}{s} \right)}
\label{eq:f}
\end{equation}
for $\phi > \phi_{\rm J}$.
The dashed lines in Figures \ref{fig:6}(a) and \ref{fig:6}(b) represent the fits obtained using data in the range $0.01 < \phi - \phi_{\rm J} < 0.06$.
The fitting parameters ($f_{0}$ and $s$) are summarized in Table \ref{table:2}.

\begin{table}
\small
  \caption{\
  Fitting parameters ($f_{0}$ and $s$) for the packing-fraction dependence of the fraction of the largest rigid cluster (Equation (\ref{eq:f})).
  The least-squares fits are performed using data in the range $0.01 < \phi - \phi_{\rm J} < 0.06$.
  }
  \begin{tabular*}{0.48\textwidth}{@{\extracolsep{\fill}}lcc}
    \hline
    Aggregate ID  & $f_{0}$ & $s$ \\
    \hline
    
    {\sf RLA-16384-1}   & $0.12$  & $0.031$  \\
    {\sf RLA-16384-2}   & $0.11$  & $0.034$  \\
    {\sf RLA-16384-3}   & $0.12$  & $0.029$  \\
    {\sf RLA-16384-4}   & $0.10$  & $0.034$  \\
    {\sf RLA-16384-5}   & $0.08$  & $0.039$  \\ \\
   
    {\sf BPCA-16384-1}  & $0.27$  & $0.037$  \\
    {\sf BPCA-16384-2}  & $0.17$  & $0.040$  \\
    {\sf BPCA-16384-3}  & $0.16$  & $0.040$  \\
    {\sf BPCA-16384-4}  & $0.21$  & $0.042$  \\
    {\sf BPCA-16384-5}  & $0.19$  & $0.038$  \\
    
    \hline
  \end{tabular*}
  \label{table:2}
\end{table}

I then calculate the coordination number within the largest rigid cluster, $Z_{\rm eff}$ (Figures \ref{fig:6}(c) and \ref{fig:6}(d)).
The largest rigid cluster is jammed at $\phi_{\rm J}$, and my numerical results show that $Z_{\rm eff} - Z_{\rm iso} \approx 0$ at the jamming point, where $Z_{\rm iso} = 4$ is the isostatic coordination number for two-dimensional aggregates.
I also find that $Z_{\rm eff} - Z_{\rm iso}$ increases linearly with $\phi - \phi_{\rm J}$ for both RLA and BPCA aggregates:
\begin{equation}
Z_{\rm eff} - Z_{\rm iso} \approx m \left( \phi - \phi_{\rm J} \right),
\end{equation}
where $m$ is the slope in Figures \ref{fig:6}(c) and \ref{fig:6}(d).
For linear repulsive systems, $Z_{\rm eff} - Z_{\rm iso}$ is known to scale as $Z_{\rm eff} - Z_{\rm iso} = m {\left( \phi - \phi_{\rm J} \right)}^{1/2}$ \citep{2003PhRvE..68a1306O}.
Thus, the linear relationship shown in Figures \ref{fig:6}(c) and \ref{fig:6}(d) is a distinctive feature of cohesive granular systems.

For aggregates of linear cohesive particles, $K$ is approximately proportional to $\phi - \phi_{\rm J}$ (Equation~(\ref{eq:bulk})).
In addition, since $\phi - \phi_{\rm J}$ increases linearly with $Z_{\rm eff} - Z_{\rm iso}$, $K$ is proportional to $Z_{\rm eff} - Z_{\rm iso}$.
This linear relationship is consistent with that found for random spring networks \citep{2009EL.....8734004E}.
\citet{2009EL.....8734004E} investigated the elastic response of two types of spring networks: one derived from real packings of linear repulsive particles, and another obtained by randomly cutting bonds in a highly connected network derived from a jammed packing.
They showed that $K$ is proportional to $Z_{\rm eff} - Z_{\rm iso}$ and remains continuous at $Z_{\rm iso}$ for random spring networks, whereas a discontinuous jump of $K$ at $Z_{\rm iso}$ appears for networks derived from repulsive particle packings.

Although both \citet{2009EL.....8734004E} and I investigate the elastic response of random spring networks, the procedures used to construct these networks differ.
As described above, they reduced the coordination number $Z$ by randomly cutting springs in the contact network obtained from a jammed packing.
In contrast, in the present study I increase $Z$ by compressing a system of cohesive particles and thereby evolving the contact network.
It is intriguing that similar critical scaling is observed in both cases, despite the fact that the evolution of $Z$ proceeds in opposite directions.

\subsection{Interparticle Compression Length}
\label{sec:delta}

\begin{figure*}[t]
\centering
  \includegraphics[width=0.8\textwidth]{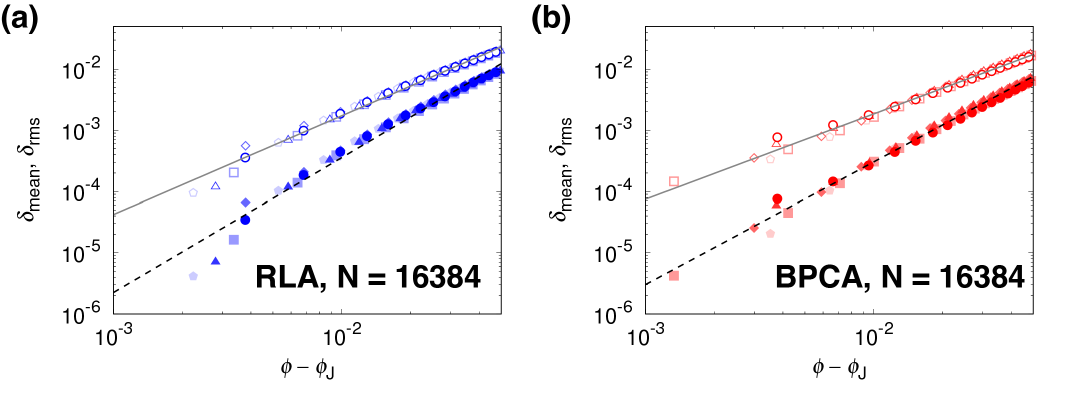}
  \caption{\
  $\delta_{\rm mean}$ and $\delta_{\rm rms}$ as functions of $\phi - \phi_{\rm J}$.
  Panels (a) and (b) show the results for the RLA and BPCA aggregates with $N = 16384$, respectively.
  Filled and open symbols correspond to $\delta_{\rm mean}$ and $\delta_{\rm rms}$, respectively.
  The dashed and solid lines represent the fits obtained via least-squares fitting (Equations (\ref{eq:delta_mean}) and (\ref{eq:delta_rms})).
  }
  \label{fig:7}
\end{figure*}

After relaxation, the force acting on a particle is determined by the elastic term (see Equation (\ref{eq:F_e})), which is proportional to the interparticle compression length.
Here, I define the mean and root-mean-square of the interparticle compression length $\delta$ as $\delta_{\rm mean}$ and $\delta_{\rm rms}$, respectively.
Figures \ref{fig:7}(a) and \ref{fig:7}(b) show $\delta_{\rm mean}$ and $\delta_{\rm rms}$ as functions of $\phi - \phi_{\rm J}$ for RLA and BPCA aggregates with $N = 16384$, respectively.
The dashed and solid lines represent the fits obtained via least-squares fitting, and I find that the following relations describe their dependence on the packing fraction:
\begin{eqnarray}
\delta_{\rm mean} & = & B {\left( \phi - \phi_{\rm J} \right)}^{\beta},  \label{eq:delta_mean} \\
\delta_{\rm rms}  & = & C {\left( \phi - \phi_{\rm J} \right)}^{\gamma}. \label{eq:delta_rms}
\end{eqnarray}
I obtain $\beta = 2.20$, $B = 9.03$, $\gamma = 1.61$, and $C = 2.90$ for the RLA case, and $\beta = 2.00$, $B = 3.06$, $\gamma = 1.39$, and $C = 1.13$ for the BPCA case.
Notably, $\beta$ is close to 2 in both cases, indicating that $\delta_{\rm mean}$ is approximately proportional to $P / k$.
In other words, the mean interparticle force $F_{\rm mean} = {( k/2 )} \delta_{\rm mean}$ scales linearly with the macroscopic pressure $P$.
This proportionality is consistent with the Rumpf equation \citep{https://doi.org/10.1002/cite.330420806}:
\begin{equation}
F_{\rm mean} \approx \frac{4 \pi P}{Z \phi},
\end{equation}
where $Z$ is the coordination number of the aggregates.
The dependence of $Z$ on the filling factor is shown in Figures \ref{fig:6}(e) and \ref{fig:6}(f).
For $\phi \approx \phi_{\rm J}$, $Z \approx 4$ and therefore $F_{\rm mean} \approx (\pi / \phi_{\rm J}) P$.

I note that Equations (\ref{eq:delta_mean}) and (\ref{eq:delta_rms}) are valid only in the regime where $\delta_{\rm mean} / \delta_{\rm rms} \ll 1$.
Figure \ref{fig:7} indicates that the ratio $\delta_{\rm mean} / \delta_{\rm rms}$ increases with $\phi$, because $\delta_{\rm mean}$ and $\delta_{\rm rms}$ are governed by different exponents, $\beta$ and $\gamma$.
However, by definition, the mean must be less than or equal to the root mean square; thus, $\delta_{\rm mean} / \delta_{\rm rms}$ should always be smaller than 1.

Since $\delta$ can take both positive and negative values, the root-mean-square value, $\delta_{\rm rms}$, characterizes the typical amplitude of $\delta$ in the system.
The scaling of $\delta_{\rm rms}$ with $\phi - \phi_{\rm J}$ (Equation (\ref{eq:delta_rms})) is therefore expected to be related to the geometric properties of the contact network, although the connection remains unclear at present.

\section{Discussion}

\subsection{Breakable Contact Model}
\label{sec:break}

In this study, I mainly investigate the compressive behavior of aggregates whose constituent particles interact via unbreakable contacts (Figure \ref{fig:1}(a)).
Although this idealized and simple setting is useful for capturing the essence of cohesive systems, the unbreakable contact model may be unrealistic in some situations.
I therefore discuss the impact of contact breakage on the compressive behavior in this section.

\begin{figure}[t]
\centering
  \includegraphics[width=\columnwidth]{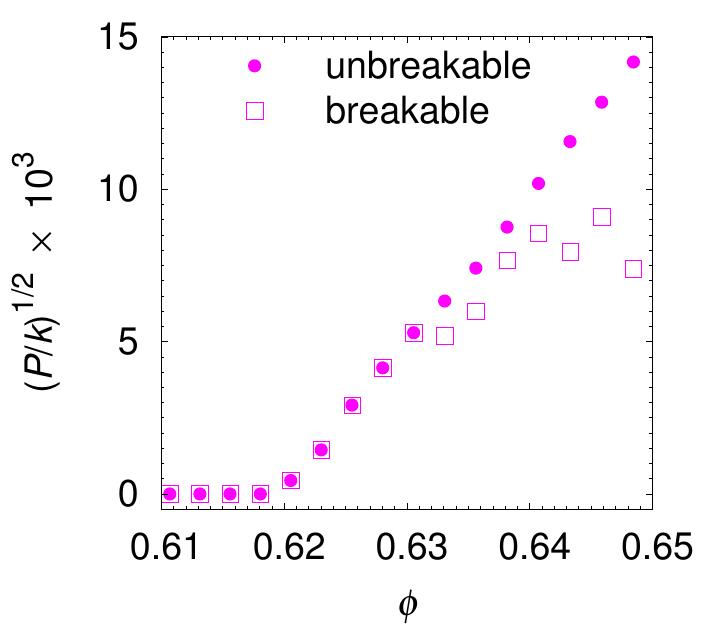}
  \caption{\
  Normalized pressure $P / k$ as a function of $\phi$.
  Filled circles indicate the results for the unbreakable contact model, whereas open squares indicate those for the breakable contact model with $\delta_{\rm c} = 0.01$.
  The pre-compression aggregate is {\sf DLA-4096-1} for both contact models.
  }
  \label{fig:8}
\end{figure}

Figure \ref{fig:8} shows the normalized pressure as a function of $\phi$.
Filled circles indicate the results for the unbreakable contact model, whereas open squares indicate those for the breakable contact model with $\delta_{\rm c} = 0.01$.
The pre-compression aggregate is {\sf DLA-4096-1} for both contact models.
It is clear that the pressure is essentially independent of the choice of contact model around the jamming point, whereas the two curves start to deviate from each other when $\phi$ exceeds a certain value (in this case, $\phi > 0.633$).
The critical scaling around the jamming point, $P/k = A {( \phi - \phi_{\rm J} )}^{2}$, is therefore a common feature of both the unbreakable and breakable contact models.

\begin{figure*}[t]
\centering
  \includegraphics[width=\textwidth]{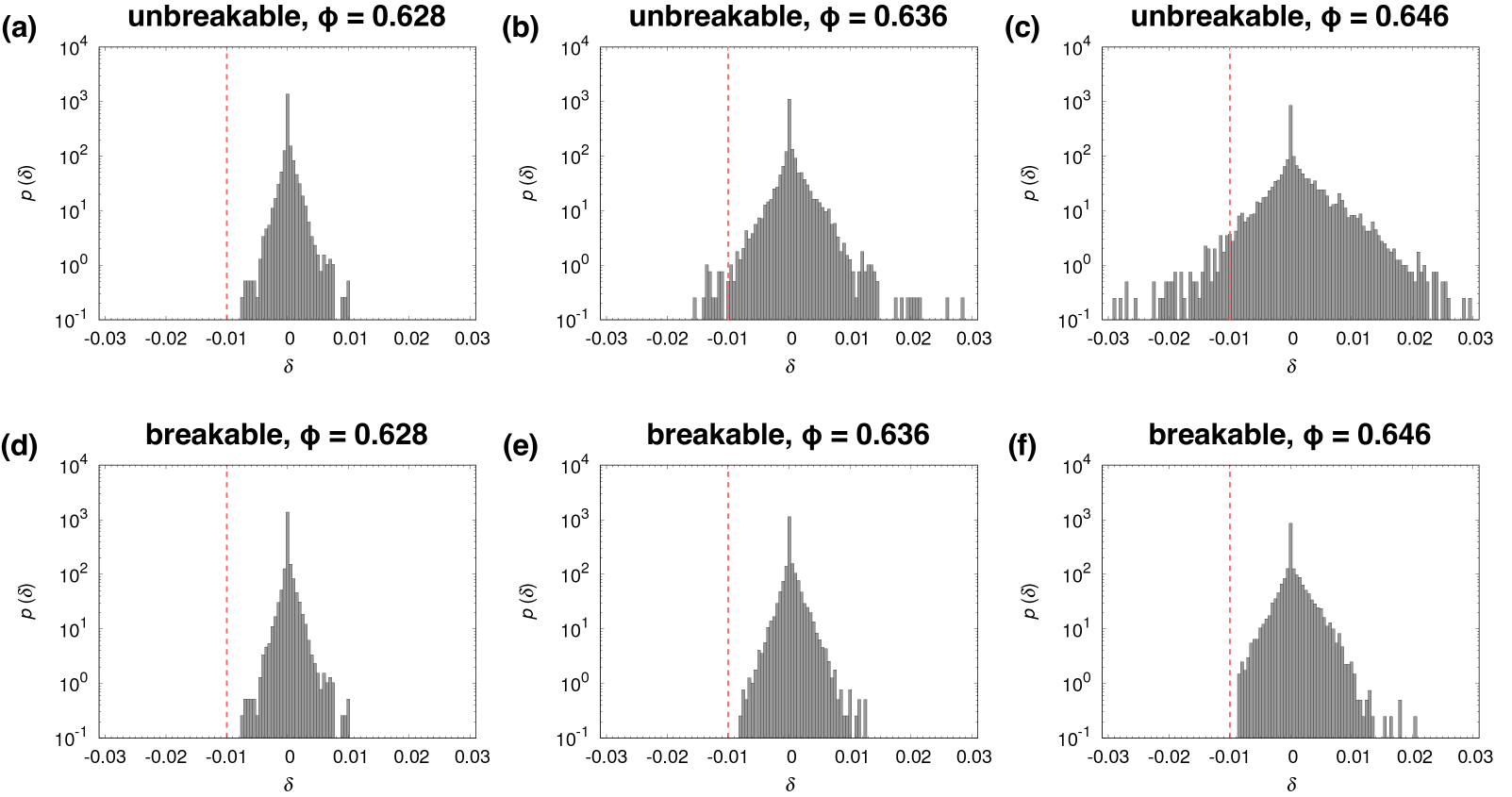}
  \caption{\
  Frequency distribution of $\delta$, $p ( \delta )$.
  Panels (a)--(c) show the results for the unbreakable contact model, and panels (d)--(f) show the results for the breakable contact model.
  Vertical dashed lines at $\delta = - \delta_{\rm c}$ indicate the critical compression length in the breakable contact model (Equation (\ref{eq:F_e_break})).
  }
  \label{fig:9}
\end{figure*}

When $\delta \ge - \delta_{\rm c}$ holds for all interparticle contacts within the aggregate, ${\bm F}_{\rm e}$ in the breakable contact model exactly coincides with that in the unbreakable contact model.
In contrast, if $\delta < -\delta_{\rm c}$ for some contacts, the values of ${\bm F}_{\rm e}$ for those contacts differ between the two models.
Figure \ref{fig:9} shows the frequency distribution of $\delta$, $p ( \delta )$.
Here, I define $p ( \delta )$ such that it satisfies the normalization condition
$\int_{- \infty}^{+ \infty} p ( \delta ) \, {\rm d}\delta = 1$.
Figures \ref{fig:9}(a)--\ref{fig:9}(c) correspond to the unbreakable contact model, and Figures \ref{fig:9}(d)--\ref{fig:9}(f) correspond to the breakable contact model.
For the unbreakable contact model, the distribution is approximately symmetric with respect to $\delta = 0$.
At $\phi = 0.628$, the distribution for the breakable contact model (Figure \ref{fig:9}(d)) is identical to that for the unbreakable contact model (Figure \ref{fig:9}(a)).
This is consistent with the fact that the pressure at $\phi = 0.628$ is independent of the contact model.

The distributions $p ( \delta )$ for the two models deviate from each other at $\phi = 0.636$ and $0.646$.
In these cases, a fraction of the contacts in the unbreakable contact model (Figures \ref{fig:9}(b) and \ref{fig:9}(c)) have $\delta < - \delta_{\rm c}$.
In contrast, in the breakable contact model (Figures \ref{fig:9}(e) and \ref{fig:9}(f)), no contacts with $\delta < - \delta_{\rm c}$ are present.
Although interparticle contacts are broken at $\delta = - 2 \delta_{\rm c}$ in the breakable contact model (Figure \ref{fig:1}(b)), contacts with $- 2 \delta_{\rm c} < \delta < - \delta_{\rm c}$ are mechanically unstable and are not expected to exist in static configurations \citep{2025JPSJ...94h4801Y}.

\subsection{Compression of Randomly Dispersed Cohesive Particles}

In this study, I investigated the compression behavior of aggregates using the DEM and determined the critical scaling of the pressure (Equation (\ref{eq:p/k})).
Although $\phi_{\rm J}$ varies depending on the initial structure, the exponent $\alpha = 2$ appears to be common.
Furthermore, my analysis of $Z_{\rm eff}$ and a comparison with random spring networks \citep{2009EL.....8734004E} suggest that this scaling reflects the properties of the interparticle interaction model (Section \ref{sec:rigid}).

\begin{figure}[t]
\centering
  \includegraphics[width=\columnwidth]{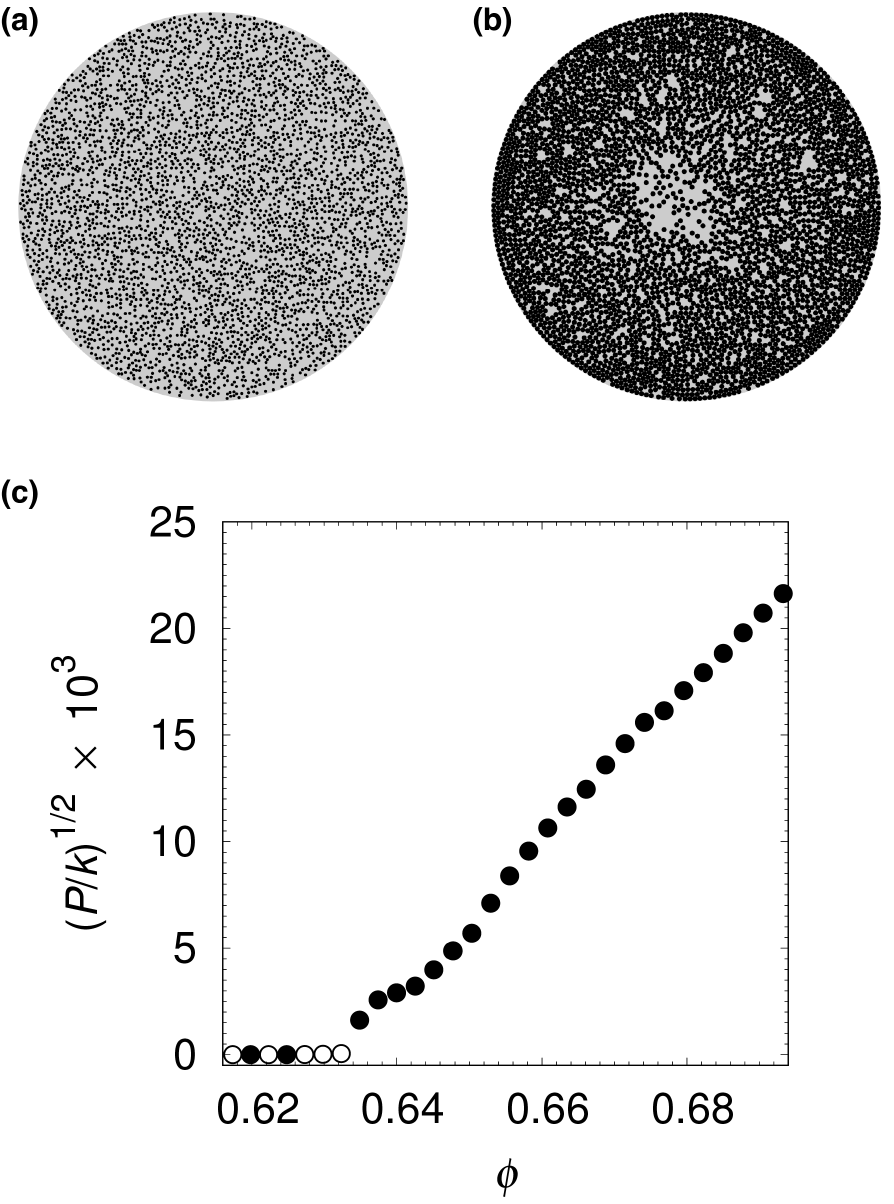}
  \caption{\
  (a) Initial configuration of randomly dispersed cohesive particles.
  (b) Snapshot of a compressed system at $\phi = 0.666$.
  (c) Normalized pressure $P / k$ as a function of $\phi$.
  Data with positive $P$ are plotted using filled symbols, whereas data with negative $P$ are plotted as $- P$ using open symbols.
  }
  \label{fig:10}
\end{figure}

This naturally raises the question: how does the compression behavior change if the particles are initially distributed randomly rather than forming aggregate structures?
To address this question, I perform additional simulations.
Using the unbreakable contact model, I carry out compression simulations of a system of randomly dispersed cohesive particles with $N = 4096$ and an initial wall radius $R_{\rm wall,ini} = 128$ (corresponding to an initial packing fraction of $0.25$).
The initial configuration is shown in Figure \ref{fig:10}(a), and the configuration at $\phi = 0.666$ is shown in Figure \ref{fig:10}(b).

Figure \ref{fig:10}(c) shows the normalized pressure $P/k$ after relaxation as a function of $\phi$.
I find that a jamming transition occurs at $\phi_{\rm J} \approx 0.63$.
As in the case of aggregates, the data near the jamming transition are approximately described by $P/k = A {\left( \phi - \phi_{\rm J} \right)}^{2}$.
This result further supports the idea that the critical scaling with exponent $\alpha = 2$ originates from random networks of linear springs.

It should be noted, however, that this result for the randomly dispersed system differs from the conclusions of earlier studies that carried out simulations under similar conditions \citep{2018PhRvL.121r8002K, 2020PhRvR...2c2047K}.
\citet{2020PhRvR...2c2047K} performed compaction simulations of randomly dispersed cohesive particles and investigated the relationship between the bulk modulus $K$ and $\phi$.
The interparticle interaction they used is such that two particles $i$ and $j$ interact when $\delta_{ij} > - 2 \delta_{\rm c}$, and the corresponding force is given by Equation (\ref{eq:F_e_break}).
In contrast to the breakable contact model adopted in the present study, they did not take into account any hysteresis between connected and separated states.
In addition, they employed periodic boundary conditions and compressed the system by repeatedly applying an affine transformation followed by relaxation.
They found a critical scaling with exponent $\alpha \approx 3.5$ and reported that $\phi_{\rm J}$ depends on $\delta_{\rm c}$.
The differences between their results and those of the present study indicate that both the exponent $\alpha$ and the jamming point $\phi_{\rm J}$ depend on the details of the interparticle attraction model and/or on the compression protocol.

\section{Summary}

In this study, I investigated the compression-driven jamming behavior of two-dimensional porous aggregates composed of cohesive, frictionless disks.
Three types of initial aggregates were prepared using different aggregation procedures (RLA, BPCA, and DLA) in order to clarify how the initial morphology affects the jamming behavior.
I showed that the system exhibits a clear jamming transition at a critical packing fraction $\phi_{\rm J}$ that depends strongly on the initial structure.
By extrapolating to infinite system size, I obtained $\phi_{\rm J} = 0.765 \pm 0.004$ for RLA aggregates, $0.727 \pm 0.004$ for BPCA aggregates, and $0.602 \pm 0.023$ for DLA aggregates.
Above the transition, the pressure follows a power-law scaling, $P = A {( \phi - \phi_{\rm J} )}^{2}$, which implies that the bulk modulus $K$ is approximately proportional to $\phi - \phi_{\rm J}$.

To understand the microscopic origin of this behavior, I performed a rigid-cluster analysis.
The jamming transition is accompanied by the percolation of the largest rigid cluster that makes multiple contacts with the confining wall.
At the jamming point, the effective coordination number within the largest rigid cluster satisfies $Z_{\rm eff} \approx Z_{\rm iso} = 4$, and for $\phi > \phi_{\rm J}$ I found that $Z_{\rm eff} - Z_{\rm iso}$ increases linearly with $\phi - \phi_{\rm J}$ (Figures \ref{fig:6}(c) and \ref{fig:6}(d)).
Consequently, the bulk modulus is proportional to $Z_{\rm eff} - Z_{\rm iso}$, in contrast to frictionless systems with purely repulsive, linear interactions, for which $Z_{\rm eff} - Z_{\rm iso}$ is known to scale as $(\phi - \phi_{\rm J})^{1/2}$ \citep{2003PhRvE..68a1306O}.
This linear trend implies that the elastic response of compressed porous aggregates is analogous to that of random spring networks \citep{2009EL.....8734004E}.

I also examined the effect of contact breakage by comparing an unbreakable contact model with a breakable contact model, in which contacts are weakened and eventually broken under finite tensile loading (Figure \ref{fig:1}(b)).
The results show that the critical scaling of the pressure near the jamming point, $P = A {( \phi - \phi_{\rm J} )}^{2}$, is essentially identical for both models, and that differences appear only at higher packing fractions where tensile contacts become unstable in the breakable model (Figure \ref{fig:8}).

In summary, this work demonstrated that the jamming landscape of cohesive soft materials is strongly modulated by the aggregate formation pathway, thereby bridging concepts from colloidal gelation, porous-media mechanics, and granular physics.
Future extensions to three-dimensional systems, frictional interactions, and shear- or tensile-driven dynamics will be invaluable for establishing a comprehensive, morphology-aware jamming framework relevant to a broad range of soft-matter physical contexts.


\section*{Conflicts of interest}

There are no conflicts to declare.

\section*{Data availability}

Data for this article are available at Zenodo (https://doi.org/10.5281/zenodo.15722246).

\section*{Acknowledgements}

This research is supported by JSPS KAKENHI Grant (JP24K17118 and JP25K01063).
Numerical computations were in part carried out on the general-purpose PC cluster at the Center for Computational Astrophysics, National Astronomical Observatory of Japan.




\renewcommand\refname{References}

\bibliography{rsc} 
\bibliographystyle{rsc} 

\end{document}